# All-angle left-handed negative refraction in Kagomé and honeycomb lattice photonic crystals

R. Gajić[1,2], R. Meisels[2], F. Kuchar[2], K. Hingerl[3]

[1] *Institute of Physics, P.O. Box 68, 11080 Belgrade, Serbia and Montenegro*

[2] *Institute of Physics, University of Leoben, Franz Josef Strasse 18, A-8700 Leoben, Austria*

[3] *Christian Doppler Lab, Institute for Semiconductors and Solid State Physics, University of Linz, Altenbergstrasse 69, A-4040 Linz, Austria*

Possibilities of all-angle left-handed negative refraction in 2D honeycomb and Kagomé lattices made of dielectric rods in air are discussed for the refractive indices 3.1 and 3.6. In contrast to triangular lattice photonic crystals made of rods in air, both the honeycomb and Kagomé lattices show all-angle left-handed negative refraction in the case of the TM2 band for low normalized frequencies. Certain advantages of the honeycomb and Kagomé structures over the triangular lattice are emphasized. This specially concerns the honeycomb lattice with its circle-like equifrequency contours where the effective indices are close to -1 for a wide range of incident angles and frequencies.

## Introduction

So far, different 2D photonic crystal (PhC) structures have been analyzed. Mostly the properties of lattices with square and hexagonal symmetry have been calculated [1-3] and have been used in the negative refraction experiments [4-9]. Derived structures have been studied together with the basic geometries. In the case of hexagonal lattice, three fundamental structures exist: the triangular, honeycomb, and Kagomé lattice (Fig. 1). The Kagomé lattice is well known in solid state physics. Magnetic materials with the Kagomé lattice and its isomers produce the most frustrated magnetic systems [10]. Regarding PhC physics, the Kagomé structure is practically neglected [11], in spite of its large gap-to-midgap ratio. On the other hand, the honeycomb or graphite lattice PhC has already been investigated [12-15] to some extent.

Recently, materials exhibiting negative refraction have attracted considerable attention based on the seminal work of Veselago [16].

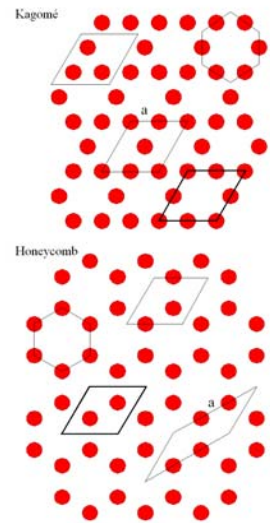

Fig. 1. The Kagomé and the honeycomb lattice are displayed with different unit cells of the same areas, $2 \cdot a^2 \sqrt{3}$ and $1.5 \cdot a^2 \sqrt{3}$, respectively. The unit cells marked with the thick line are used in the calculations.

These structures are periodic and can be either realized with so-called, metamaterials [17-19] or photonic crystals. Depending on the product sign of $v_{ph} \cdot v_{gr}$, we distinguish the right-handed (RH⁻) and left-handed (LH⁻) negative refraction [20] ($v_{ph}$ and $v_{gr}$ are the phase phase and group velocities in PhC). RH⁻ refraction can be realized in PhCs for round equifrequency contours (EFC) with inward gradients. This takes place around symmetrical points in the Brillouin zone, away from the Γ point, i.e. in the vicinity of the M point in 2D square lattice PhCs. In this case, Luo et al. [21] proposed all-angle negative refraction (AANR) for the GaAs dielectric contrast. The LH⁻ refraction is also possible in PhCs in higher bands around the Γ point [22,23]. There, $v_{ph}$ and $v_{gr}$ can be anti-parallel like in Veselago's metamaterials. One of the tasks is to search for PhC structures with all-angle left-handed negative refraction (AALNR), particularly with all-angle Veselago negative refraction where $n_{eff}$ = -1, although PhCs with $n_{eff}$ = -1 and metamaterials with $n_{ph}$ = -1 can have different optical properties [24]. In this work we are looking for an all-angle $n_{eff}$ = -1 candidate in the hexagonal PhC family due to their circular EFCs. We will use two different effective indices of refraction, $n_{beam}(\theta_{in},\omega) = \sin(\theta_{in})/\sin(\theta_r)$ and $n_{peff}(\theta_{in},\omega) = \text{sgn}(v_{gr} \cdot k_{PhC}) \cdot c |k_{PhC}|/\omega \equiv \pm k_{PhC}/k_{air}$. The former is obtained from a Snell-like formula [4,5] and represents an effective index of refraction in the $k_{in}$ direction, and the latter is an effective phase index of refraction. The signs of $n_{beam}$ and $n_{peff}$ reflect the positive or negative refraction, and the right-handed or left-handed behavior in PhCs, respectively. In the case of RH⁻ or LH⁺ refraction [20], $n_{beam}$ and $n_{peff}$ have different signs.

## Calculation

Here, we study Kagomé and honeycomb dielectric rod PhCs and their TM/TE bands where LH⁻ refraction takes place (electric and magnetic fields are parallel to the rods for TM and TE modes, respectively). We start with a 2D Kagomé PhC made of dielectric rods in air and analyze the EFC and propagation properties. As a tool the plane-wave method (PWM) with BandSOLVE [25] and the finite-difference time-domain (FDTD) calculation with FullWAVE [26] are used. At the beginning, the band structure and the gap-map are determined.

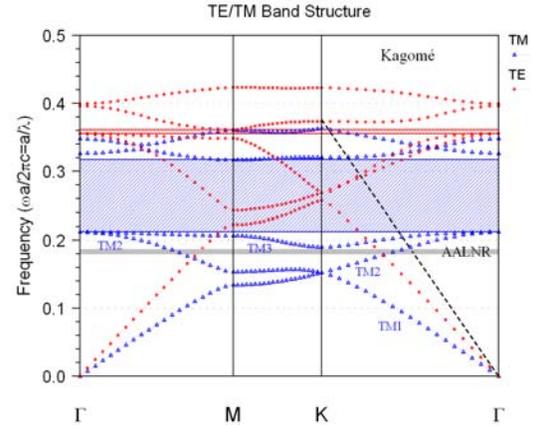

Fig. 2. The band structure of a 2D Kagomé GaAs rod photonic crystal for $r/a \simeq 0.27$. The dashed line corresponds to the light line and the gray strip denotes the region with AALNR in the TM2 band.

The calculations are performed for two different dielectric rod materials ($n$ = 3.1 and $n$ = 3.6). At microwave frequencies, $n$ = 3.1 and 3.6 can be realized using $Al_2O_3$ (alumina) and GaAs as the rod material. The Kagomé lattice has a large gap-to-midgap ratio $\Delta\omega/\omega_o$ (an omnidirectional band gap) which is of the same order as in the triangular lattice. Namely, $\Delta\omega/\omega_o \geq 35$ % and $\geq 40$ % for $n$ = 3.1 and $n$ = 3.6, respectively. In a GaAs rod PhC AALNR is studied for $r/a$

≃ 0.27 ($r$ is the rod radius and $a$ the smallest distance between the rods). The Kagomé unit cell has a basis with three rods per cell, as shown in Fig. 1. The band calculations, based on PWM, for the first 5 TM/TE bands in the case of the GaAs rods, are presented in Fig. 2. For $n = 3.6$, $r$ and $a$ are 0.6 mm and 2.2 mm. A possible candidate for AALNR is the TM2 band whose EFCs are given in Fig. 3.

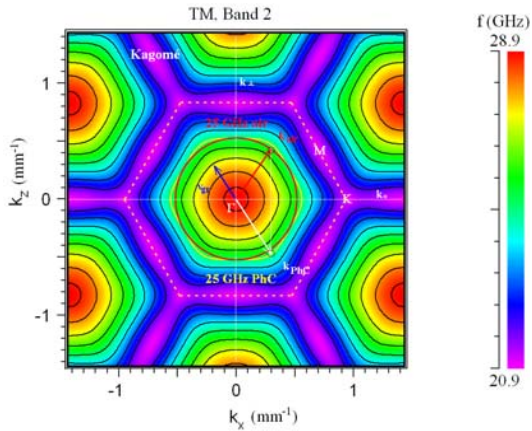

Fig. 3. The equifrequency contours of the TM2 mode for a 2D Kagomé GaAs rod photonic crystal. The red circle denotes the air 25 GHz EFC ($\tilde{f} \simeq 0.183$) whereas, the yellow thick contour is a PhC EFC at the same frequency. EFCs are presented for 22, 23, … , 28 and 28.7 GHz. The red arrow denotes the direction of the incident wave across the ΓK interface at 34°, the white one is for $k_{PhC}$ and the blue arrow stands for the group velocity, $v_{gr}$. $k_\parallel$ and $k_\perp$ are $k$-vectors parallel and normal to the ΓK interface, respectively. The dashed line determines the first Brillouin zone.

The equifrequency contours are shown in the microwave range between 22 and 28.7 GHz but the conclusions are, of course, quite general due to the scaling property of PhCs. The necessary conditions for AALNR are a convex EFC in the PhC around Γ with inward gradients and such a PhC EFC that surrounds the corresponding one in air. Additionally, $\lambda = c/f \geq 2 \cdot a_s$ should be fulfilled ($a_s$ is the surface lattice constant, $a_s^K = a$ for the ΓK interface of the Kagomé lattice) in order to avoid diffraction [23]. The EFC structure in Fig. 3 reveals that the air circles are of comparable size with those in the PhC. For the $n = 3.6:1$ rod contrast, AALNR takes place for the frequencies between 24.5 and 25.3 GHz ($\tilde{f} \simeq 0.18$ to 0.186) as it is marked in Fig. 2. As an example, we analyze AALNR near the upper bound, at 25 GHz (or the normalized frequency, $\tilde{f} = fa/c \simeq 0.183$). The air EFC is inside the corresponding one in the PhC as in Fig. 3. In our case, for 25 GHz, the condition regarding the lack of diffractions also holds ($\lambda > 5.4 \cdot a_s^K$). Moreover, for the incident angles, $\theta_{in} > 35°$ $v_{ph}$ and $v_{gr}$ are nearly anti-parallel like in Veselago metamaterials. The wave pattern of the 25 GHz TM2 electromagnetic wave (EMW) propagating through the Kagomé lattice with $n_{beam} \simeq -1$ is presented in Fig. 4.

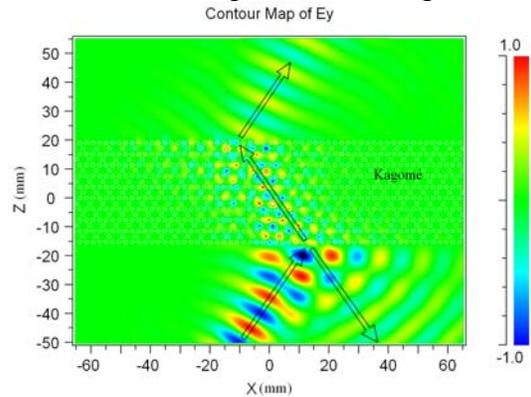

Fig. 4. The wave pattern of the 25 GHz TM2 EMW propagating through a Kagomé GaAs rod photonic crystal at the 34° incidence across the ΓK interface (FDTD).

A careful FDTD analysis revealed backward waves in PhC. AALNR appears for the TM2 band positioned below the first gap in a Kagomé-lattice PhC as in Fig. 2. As the dielectric contrast increases, the gap normally widens and the mid-gap moves to lower frequencies enabling engineering until

GAJIC et al.

the AALNR condition is fulfilled. AALNR is found for the $n = 3.1$ rods too but in the narrow frequency range, 0.4 GHz wide.

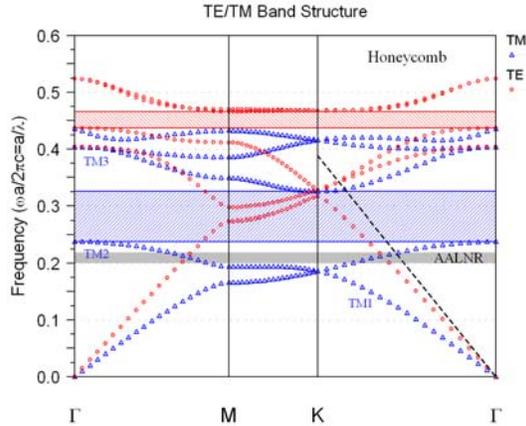

Fig. 5. Band structure of a 2D honeycomb GaAs rod photonic crystal for $r/a \simeq 0.24$. The dashed line corresponds to the light line and the gray strip represents the region with AALNR in the TM2 band.

It is worth emphasizing that LH⁻ refraction is also present for the TM3 band but only for small incident angles, since the EFCs are star-like.

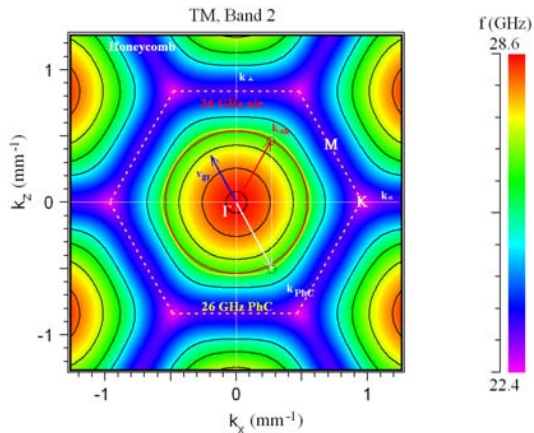

Fig. 6. The equifrequency contours of the TM2 mode for a 2D honeycomb GaAs rod photonic crystal. The red circle stands for the air 26 GHz EFC ($f\,\tilde{}\, \simeq 0.22$) whereas, the yellow contour corresponds to an EFC of the same frequency in the crystal. The EFCs are presented for 25, 26, ... , 28 and 28.5 GHz. The red arrow indicates the incident wave at 30°. The rest is the same as in Fig. 3.

We find the similar situations for the TE EFCs bands, which are not circular. This concerns structures both with rods and with holes. TE gaps are smaller in dielectric rod PhCs since as a rule, isolated high-dielectric regions favour the TM band gaps [12]. The effective indices $n_{beam}$ and $n_{peff}$ of the 25 GHz TM2 beam are shown in Fig. 8. $n_{beam}$ becomes large for small incident angles but always negative.

In the honeycomb lattice the adjacent hexagons share two rods instead of one rod as in the Kagomé structure (see Fig. 1). The honeycomb unit cell has a basis with two rods per cell and it is smaller than the Kagomé one. For the $n = 3.6$ rods, the honeycomb lattice has a lower gap-to-midgap ratio than Kagomé PhC, $(\Delta\omega/\omega_o)_{TM} \simeq 31\ \%$.

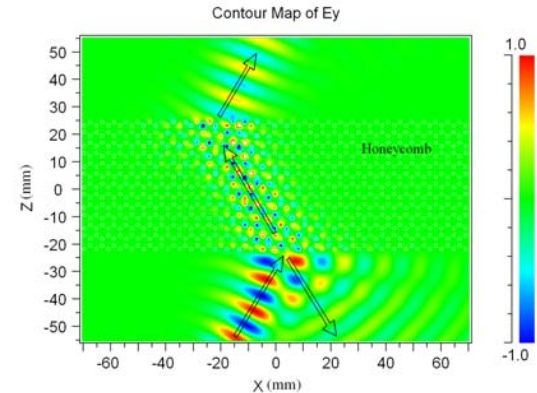

Fig. 7. The wave pattern of the 26 GHz TM2 EMW propagating through a honeycomb GaAs rod photonic crystal at the 30° incidence across the ΓK interface (FDTD).

Comparing the band structures in Figs. 2 and 5, there is an important difference between the two lattices regarding the TM bands with negative refraction. In the Kagomé lattice the TM2 and TM3 bands overlap in the Γ point whereas, they are separated by the band gap in the honeycomb PhC, which is favourable for applications. Here, we study the structure for $r/a \simeq 0.24$ ($r = 0.6$ mm). Figs. 6 and 7 show the TM2 EFCs and the corresponding wave propagation at 30° in a honeycomb GaAs rod lattice. Similarly, to the Kagomé lattice, the EFCs are round and can be of the same size as the corresponding air EFCs. For example, the air 26 GHz TM2 EFC ($\tilde{f} \simeq 0.22$) is located within the PhC EFC of the same frequency giving rise AALNR. Again the single beam behavior of the 26 GHz wave is insured ($\lambda \simeq 2.7 \cdot a_s^H$, $a_s^H = a\sqrt{3}$ for the ΓK interface). Like in the Kagomé PhC, LH$^-$ refraction without AALNR is present in higher bands (TE2, TM3) of the honeycomb lattice PhC. The effective indices of the TM2 band are calculated and presented in Fig. 8. The honeycomb lattice is more compact than the Kagomé one, its EFCs are generally more round as the angle dependence of $n_{beam}$, and $n_{peff}$ shows in Fig. 8.

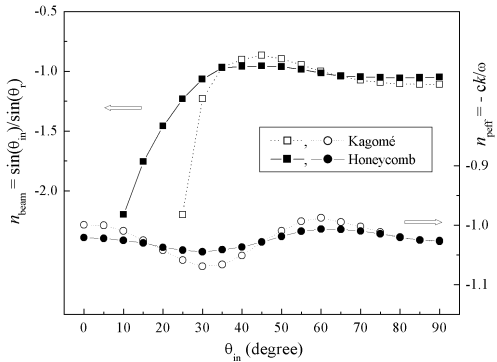

Fig. 8 (color online) The effective indices, $n_{beam}$ and $n_{peff}$ of both lattices at 25 GHz (Kagomé) and 26 GHz (honeycomb) as a function of the incident angle.

In addition, the effective indices are quite close to -1 for a wide range of incident angles in the honeycomb PhCs. For example, at 26 GHz, $-1.06 \leq n_{beam} \leq -0.96$ for $\theta_{in} \geq 30°$, and $-1.04 \leq n_{peff} \leq -1.00$ for $0 \leq \theta_{in} \leq 90°$ as in Fig. 8. Also, AALNR is present in the honeycomb PhCs made of alumina rods in air for the TM2 band. For the $Al_2O_3$ rod PhCs, the EFCs are just slightly less round comparing to the EFCs of GaAs rod PhCs. Our calculations of a triangular lattice made of $Al_2O_3$ does not show AALNR. Comparing to the Kagomé lattice, the honeycomb structure has an additional advantage since the frequency range where AALNR takes place are larger. In the case of the 3.6:1 dielectric contrast, AALNR ranges from 24.1 to 26.1 GHz ($\tilde{f} \simeq 0.2$ to 0.218). This makes $(\Delta f/f_m)_{AALNR} \approx 8\%$ ($\Delta f$ is the frequency bandwidth and, $f_m$ is the mid-frequency of the AALNR range). The similar ratio holds for the 3.1:1 contrast.

Notomi [22] investigated a 2D triangular GaAs rod photonic crystal with the ratio, $r/a = 0.35$. The effective index of the TE modes was negative ($-0.7 < n_{peff} < 0$) and well defined in the frequency range, $0.59 < \tilde{f} < 0.635$. Additionally, he demonstrated negative refraction of the TM modes in a 2D triangular GaAs air holes PhC ($r/a = 0.4$) in the range, $0.3 \leq \tilde{f} \leq 0.35$. The $n_{peff}$ of the TM band varied between -1.2 and 0. Later many authors exploited this structure investigating negative refraction and superlensing e.g. [27]. In this communication, we demonstrate for the first time possibilities of all-angle left-handed negative refraction in dielectric rod PhCs with the Kagome and the honeycomb lattice. We find that the honeycomb PhC can be particularly useful with its round equifrequency contours and the effective indices close to -1 for a wide range of incident angles. Both lattices show AALNR even for the lower dielectric contrast of $Al_2O_3$/air.

## Summary

In review, all-angle left-handed negative refraction in 2D Kagomé and honeycomb lattices made of rods in air is demonstrated. The EFC calculations show that AALNR is present for the TM2 band for both photonic crystals at low normalized frequencies. Both structures exhibit AALNR for the contrast, $n = 3.1:1$. The honeycomb lattice seems to be particularly advantageous. The effective indices, $n_{beam}$ and $n_{peff}$ are close to -1 for a wide range of incident angles and moderate dielectric contrasts and, with a useful AALNR frequency range, $(\Delta f/f_m)_{AALNR} \approx 8\%$. The Kagomé lattice PhC has less round EFCs and the smaller AALNR frequency range (around 3 % for $n = 3.6:1$ and, just 1 % for $n = 3.1:1$). Additionally, for both lattices AALNR is present for low normalized frequencies, $\tilde{f} \approx 0.2$ which eliminates undesired diffraction. Previously, AALNR with the effective indices close -1 has exclusively been related to triangular lattice PhCs of air holes in high dielectric constant materials. Taking into account that fabrication of 2D photonic crystals made of dielectric rods is sometimes a simpler task than those of air holes, here we suggest the new PhC structures of dielectric rods in air that are well suited to exhibit AALNR at lower normalized frequencies, and dielectric contrasts.

## Aknowledgments

R. G. acknowledges support by Serbian Ministry of Science and Environment Protection. The Christian Doppler Laboratory is grateful to Photeon and Dr. Heinz Syringer from Photeon Technologies for financial support and to Dr. Johann Messner from the Linz Supercomputer Center for technical support.